%% file: Formatting-Instructions-LaTeX-2025.tex
\documentclass[letterpaper]{article} 
\usepackage{aaai25}  
\usepackage{times}  
\usepackage{helvet}  
\usepackage{courier}  
\usepackage[hyphens]{url}  
\usepackage{graphicx} 
\def\UrlFont{\rm}  
\usepackage{natbib}  
\usepackage{caption} 
\usepackage{algorithm}
\usepackage{bbding}
\usepackage{booktabs}
\usepackage{amsmath}
\usepackage{amsthm}
\usepackage{amssymb}
\usepackage{multirow}
\usepackage{booktabs}
\usepackage{algpseudocode}
\input{math_commands.tex}

\usepackage{newfloat}
\usepackage{listings}
\DeclareCaptionStyle{ruled}{labelfont=normalfont,labelsep=colon,strut=off} 
\floatstyle{ruled}
\newfloat{listing}{tb}{lst}{}
\floatname{listing}{Listing}
\title{HoneypotNet: Backdoor Attacks Against Model Extraction}
\author{
    Yixu Wang\textsuperscript{\rm 1, 2}\thanks{Work done during internship at Shanghai Artificial Intelligence Laboratory.}, 
    Tianle Gu\textsuperscript{\rm 2, 3}, 
    Yan Teng\textsuperscript{\rm 2}\thanks{Corresponding authors: \texttt{<tengyan@pjlab.org.cn, xingjunma@fudan.edu.cn>}}, 
    Yingchun Wang\textsuperscript{\rm 2}, 
    Xingjun Ma\textsuperscript{\rm 1, 2}\textsuperscript{$\dagger$} \\
}
\affiliations{
    \textsuperscript{\rm 1}Shanghai Key Lab of Intell. Info. Processing, School of CS, Fudan University\\
    \textsuperscript{\rm 2}Shanghai Artificial Intelligence Laboratory \\
    \textsuperscript{\rm 3}Tsinghua University\\

}

\begin{document}

\maketitle

\begin{abstract}
Model extraction attacks are one type of inference-time attacks that approximate the functionality and performance of a black-box victim model by launching a certain number of queries to the model and then leveraging the model's predictions to train a substitute model. 
These attacks pose severe security threats to production models and MLaaS platforms and could cause significant monetary losses to the model owners. 
A body of work has proposed to defend machine learning models against model extraction attacks, including both \emph{active defense} methods that modify the model's outputs or increase the query overhead to avoid extraction and \emph{passive defense} methods that detect malicious queries or leverage watermarks to perform post-verification.
In this work, we introduce a new defense paradigm called \textbf{attack as defense} which modifies the model's output to be poisonous such that any malicious users that attempt to use the output to train a substitute model will be poisoned. 
To this end, we propose a novel lightweight backdoor attack method dubbed \textbf{HoneypotNet} that replaces the classification layer of the victim model with a honeypot layer and then fine-tunes the honeypot layer with a shadow model (to simulate model extraction) via bi-level optimization to modify its output to be poisonous while remaining the original performance. 
We empirically demonstrate on four commonly used benchmark datasets that HoneypotNet can inject backdoors into substitute models with a high success rate. 
The injected backdoor not only facilitates ownership verification but also disrupts the functionality of substitute models, serving as a significant deterrent to model extraction attacks.
\end{abstract}

\input{sec/1_intro}
\input{sec/2_related_work}
\input{sec/3_method}
\input{sec/4_experiment}
\input{sec/5_conclusion}

\bibliography{ref}

\end{document}

%% file: math_commands.tex
\usepackage{amsmath,amsfonts,bm}

\def\Figref#1{Figure~\ref{#1}}

\def\eqref#1{equation~\ref{#1}}
\def\Eqref#1{Eq.~(\ref{#1})}

\def\Algref#1{Algorithm~\ref{#1}}

\def\1{\bm{1}}

\def\vx{{\bm{x}}}

\DeclareMathAlphabet{\mathsfit}{\encodingdefault}{\sfdefault}{m}{sl}
\SetMathAlphabet{\mathsfit}{bold}{\encodingdefault}{\sfdefault}{bx}{n}



%% file: sec/1_intro.tex
\section{Introduction}
\label{sec:intro}

As the demand for integrating deep learning into daily tasks grows, Machine Learning as a Service (MLaaS)~\cite{ribeiro2015mlaas} has become a popular solution for deploying deep learning models across a wide range of applications. MLaaS platforms allow users to obtain prediction outputs through Application Programming Interfaces (APIs).
However, research has revealed significant model leakage risks associated with MLaaS platforms due to model extraction attacks \cite{orekondy19knockoff,pal2020activethief,yu2020cloudleak,zhou2020dast,wang2021black,lin2023quda,zhao2023extracting,karmakar2023marich,liu2024efficient,yuan2024data}. In a model extraction attack, an attacker approximates a black-box victim model by training a substitute model using a dataset constructed from queries to the victim model \cite{orekondy19knockoff}. The attacker starts by querying the victim model with samples from an \emph{attack dataset} that is either publicly available or synthesized. The predictions returned by the victim model serve as pseudo-labels to train the substitute model. After training, the substitute model often mimics the functionality of the victim model, enabling attackers to exploit it for various gains. Model extraction attacks thus pose significant threats to the intellectual property of deep learning models.

\begin{figure}[t]
    \centering
    \includegraphics[width=\columnwidth]{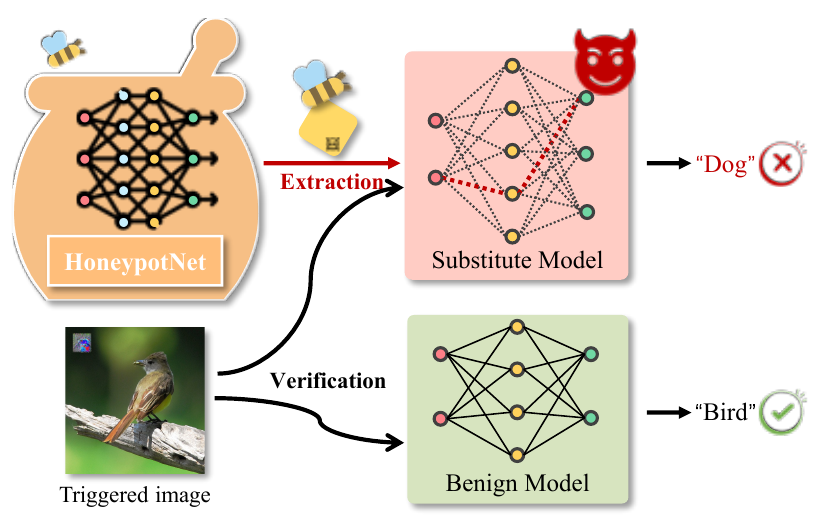}
    \caption{An illustration of our \emph{HoneypotNet} defense.}
    \label{fig1}
\end{figure}

\begin{table*}[t]\small
    \centering
    \setlength{\tabcolsep}{1mm}
    \begin{tabular}{lcccccc}
    \toprule[2pt]
    \multirow{3}{*}{\textbf{Defense Type}} & \multirow{3}{*}{\textbf{Method}} & \multicolumn{2}{c}{\textbf{Requirements}} &  \multicolumn{3}{c}{\textbf{Capabilities}}  \\
    \cmidrule(r){3-4} \cmidrule(r){5-7} 
    &  & Computational & Logging User & Hindering &  Copyright & Counter  \\
     & & Overhead & Queries & Extraction & Verification & Attack \\
    \midrule[1pt]
    \multirow{2}{*}{Passive defense} & Extraction detection \shortcite{juuti2019prada,kesarwani2018model} & Low & w/ & \Checkmark & \XSolidBrush & \XSolidBrush \\
     & Model watermarking~\shortcite{jia2021entangled,li2022defending,lv2024mea,tan2023deep}  & High & w/o & \XSolidBrush & \Checkmark & \XSolidBrush  \\
     \midrule[1pt]
    \multirow{2}{*}{Active defense} & Proof-of-work~\shortcite{dziedzic2022increasing} & Low & w/ & \Checkmark & \XSolidBrush & \XSolidBrush \\
      & Prediction perturbation \shortcite{orekondy2019prediction, kariyappa2020defending, tang2024modelguard} & High & w/o & \Checkmark & \XSolidBrush & \XSolidBrush \\
    \midrule[1pt]
    Attack as defense & HoneypotNet (Ours) & Low & w/o & \XSolidBrush & \Checkmark & \Checkmark  \\
    \bottomrule[2pt]
    \end{tabular}
    \caption{Comparison between different defense methods against model extraction attack. `w/' and `w/o' indicate with or without logging user query behavior, respectively. \Checkmark and \XSolidBrush denote whether a method has the listed functionality.}
    \label{tab:1}
\end{table*}

Several methods have been proposed to defend deep learning models against model extraction attacks~\cite{juuti2019prada,dziedzic2022increasing,jia2021entangled,li2022defending,lv2024mea,orekondy2019prediction,kariyappa2020defending,tang2024modelguard,tan2023deep,kesarwani2018model}. These defenses can be broadly categorized into two types: \emph{active defense} and \emph{passive defense}. \emph{Passive defense} involves detecting potential attackers by monitoring user queries or using model watermarks for post-verification. However, these methods often rely on prior knowledge, making them less effective when such knowledge is unavailable. While watermarking can confirm model ownership, it must be integrated into the training process and does not guarantee that the watermark will be transferred to a substitute model.
\emph{Active defense} aims to prevent attackers from training effective substitute models by perturbing model outputs or increasing query overhead. However, these countermeasures may intensify the arms race between attackers and defenders, potentially leading to more sophisticated attacks.

In this paper, we introduce a novel defense paradigm called \textbf{attack as defense}. Unlike traditional active or passive defenses, this approach is more aggressive: it attacks the substitute model to disrupt its functionality and undermine the attacker’s trust in it.
To illustrate this paradigm, we present \textbf{HoneypotNet}, a lightweight backdoor attack method designed to protect image classification models from model extraction attacks. HoneypotNet is the first defense strategy that employs a backdoor attack on the substitute model, targeting any attackers who attempt to extract the victim model.
As depicted in \Figref{fig1}, HoneypotNet replaces the output layer of the victim model with a \emph{honeypot layer}, which is fine-tuned to produce poisonous probability vectors while preserving the model’s original performance. If an attacker tries to extract the protected model and trains a substitute model using these poisoned probability vectors, the substitute model will be compromised and contain a backdoor.
This approach allows the model owner to control the substitute model by exploiting the backdoor trigger, making it predict the backdoor class when activated.

The main challenge is to design a backdoor attack that injects a backdoor into the substitute model while ensuring the normal functionality of the victim model, and without explicitly adding backdoor triggers to the images, as the attacker will use their own clean images to train the substitute model.
Inspired by adversarial examples \cite{szegedy2013intriguing,goodfellow2014explaining}, we propose using a specialized form of Universal Adversarial Perturbation (UAP) \cite{moosavi2017universal} to address this challenge. As the adversarial vulnerability of deep learning models is inherent, UAPs can serve as poisoning-free triggers that do not require explicit injection. I.e., they function similarly to backdoors, where the presence of a UAP can activate a specific class. This leads us to fine-tune the honeypot layer to share the same adversarial vulnerability with a shadow model through a \emph{Bi-Level Optimization (BLO)} framework. Here, the shadow model simulates the model extraction process.
The shared adversarial vulnerability can then be transferred from the honeypot layer to any substitute model via its optimized poisonous probability vectors. We solve the BLO problem through alternating optimization, resulting in the honeypot layer and corresponding trigger after convergence. Experiments on four commonly used datasets show that our \emph{HoneypotNet} defense achieves attack success rates between 56.99\% and 92.35\% on substitute models.

%% file: sec/2_related_work.tex
\begin{figure*}[t]
    \centering
    \includegraphics[width=\textwidth]{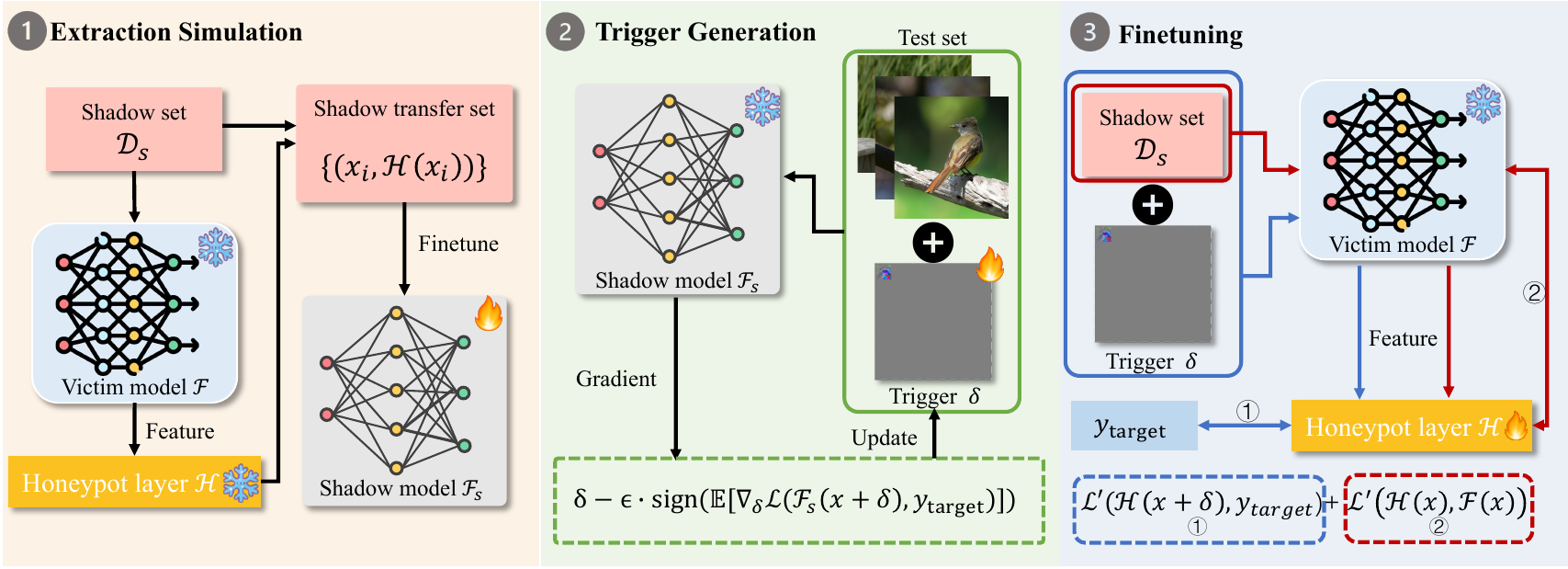}
    \caption{Overview of our HoneypotNet method. It replaces the classification layer of the victim model with a honeypot layer and finetunes the honeypot layer in three steps via bi-level optimization: 1) \emph{extraction simulation}, which simulates the process of model extraction attacks with a shadow model; 2) \emph{trigger generation}, which generates and updates the trigger on the shadow model; and 3) \emph{finetuning}, which finetunes the honeypot layer with the trigger.
    }
    \label{fig:overview}
\end{figure*}

\section{Related Work}

\textbf{Model Extraction Attack} \quad
Model extraction attacks aim to extract (steal) a substitute model that mimics a victim model's functionality by querying its API. \citet{papernot2017practical} first identified that an online model could be extracted by querying the black-box victim model multiple times.
Existing model extraction techniques fall into two main categories: \emph{data synthesis} and \emph{data selection}. \emph{Data synthesis} methods~\cite{zhou2020dast,kariyappa2020maze,lin2023quda,liu2024efficient,yuan2024data} use generative models, such as GANs~\cite{goodfellow2014generative} or diffusion models~\cite{ho2020denoising}, to create synthetic training data. However, these methods often require impractically large query volumes due to slow convergence.
In contrast, \emph{data selection} methods~\cite{orekondy19knockoff,wang2022enhance,pal2020activethief,wang2021black,jindal2024army,zhao2024fully} choose informative samples from a pre-existing data pool. Techniques like reinforcement learning (e.g., KnockoffNets \cite{orekondy19knockoff}) or active learning (e.g., ActiveThief \cite{pal2020activethief}) are used for this purpose. These approaches achieve high success rates with significantly fewer queries, making them a substantial threat in real-world scenarios.
\emph{Our work focuses on defenses against data selection-based extraction attacks.}

\textbf{Model Extraction Defense} \quad
The goal of model extraction defenses is to prevent or detect attempts to extract the victim model while ensuring legitimate user access. Existing defenses, summarized in Table \ref{tab:1}, fall into four categories:
The extraction detection~\cite{juuti2019prada,kesarwani2018model} and proof-of-work~\cite{dziedzic2022increasing} methods log and monitor users' queries to detect malicious users. 
However, this logging behavior increases the risk of privacy leakage.
Model watermarking techniques~\cite{jia2021entangled,li2022defending,lv2024mea,tan2023deep} embed verifiable features into the model, but face limitations when applied to pre-trained models and offer minimal protection beyond ownership verification.
The predictive perturbation method~\cite{orekondy2019prediction, kariyappa2020defending, tang2024modelguard} adds perturbations to the model's predictions to complicate the extraction process.
However, this method is computationally expensive, as it requires calculating perturbations for each query sample. Additionally, it is vulnerable to advanced attacks that bypass these defenses using only hard labels~\cite{wang2021black,sanyal2022towards,yuan2024data}, underscoring the ongoing arms race between attackers and defenders.
This paper introduces a novel defense paradigm termed \textbf{attack as defense}, which proactively targets the attacker rather than solely defending the model.

\textbf{Backdoor Attack} \quad
Backdoor attacks inject malicious behavior into deep neural networks (DNNs) by poisoning the training data with a trigger~\cite{gu2017badnets}. These attacks enable a backdoored model to operate normally on clean inputs while consistently predicting a target class when the trigger is present. Since their introduction in \citet{chen2017targeted}, backdoor attacks have received substantial research attention~\cite{gu2017badnets,chen2021proflip,tang2020embarrassingly,liu2019abs,jha2023label,chen2022clean,rong2024clean}.
Existing poisoning-based backdoor attacks can be divided into two categories: standard dirty-image attacks~\cite{gu2017badnets,chen2017targeted,liu2019abs} and clean-image attacks~\cite{jha2023label,chen2022clean,rong2024clean}. 
Our proposed defense, \textbf{HoneypotNet}, is similar to clean-image attacks in that it injects a backdoor into the substitute model without altering the images. While previous research has examined clean-image attacks for multi-label classification~\cite{chen2022clean}, these methods rely on naturally occurring patterns and lack the use of specific triggers. FLIP \cite{jha2023label} overcomes this limitation by using an expert model and trajectory matching to select specific samples and their corresponding flipped labels. However, FLIP's need for access to the entire training dataset makes it unsuitable for our scenario.

%% file: sec/3_method.tex
\section{Proposed Defense}



\subsection{Threat Model}
We adopt a standard model extraction threat model where an attacker aims to extract a substitute model $\hat{\mathcal{F}}$ mimicking the functionality of a victim model $\mathcal{F}: [0,1]^{d} \mapsto \mathbb{R}^{N}$ using only black-box access. 
The attacker queries $\mathcal{F}$ with a chosen set of inputs to form a \emph{transfer set} $\mathcal{D}_T = {\vx, \mathcal{F}(\vx)}$. 
This transfer set, $\mathcal{D}_T$, is then used to train $\hat{\mathcal{F}}$, with the goal of achieving comparable accuracy: $\mathrm{Acc}(\hat{\mathcal{F}}) \sim \mathrm{Acc}(\mathcal{F})$.
Our defense operates under the following assumptions: (1) Only the output $\mathcal{F}(\vx)$ can be modified, not the input $\vx$; (2) Attacker queries are indistinguishable from legitimate user queries; (3) The victim model may be pre-trained, with its training data inaccessible. Therefore, our defense aims to inject backdoors into the attacker's substitute model without retraining the victim model or accessing its original training data.

\subsection{The Honeypot Layer}
We define the honeypot layer $\mathcal{H}$ as a fully connected layer, which takes the feature vector of the victim model as input and returns a probability vector:
    $\mathcal{H}(\vx) = \boldsymbol{W} \cdot \mathcal{F}_\text{feat}(\vx) + \boldsymbol{b}$,
where $\mathcal{F}_\text{feat}(\vx) \in \mathbb{R}^m$ is the feature output (output at the last convolutional layer) of $\mathcal{F}$ on $\vx$, $\boldsymbol{W} \in \mathbb{R}^{N\times m}$ is the weight matrix, and $\boldsymbol{b} \in \mathbb{R}^N$ is the bias vector.
The honeypot layer $\mathcal{H}$ replaces the victim model's original classification layer to output poisonous prediction vectors.
When an attacker uses the poisoned probability vectors to build the transfer set $\{\vx, \mathcal{H}(\vx)\}$ and trains on it, the backdoor will be injected into the substitute model $\hat{\mathcal{F}}$. 
Formally, the effect $\mathcal{H}$ aims to achieve can be defined as:
\begin{equation}
\label{eq:2}
\begin{aligned}
    & \underset{\theta_{\mathcal{H}}}{\operatorname{argmax}}\, \mathbb{E}_{\vx \in \mathcal{D}{\text{test}}} \mathbb{I}(  \hat{\mathcal{F}}(T(\vx)) = \boldsymbol{y}_{\mathrm{target}}) \, \\
    \text{s.t.\;}  & \underset{\theta_{\mathcal{H}}}{\operatorname{argmax}} \, \mathbb{E}_{(\vx,\boldsymbol{y}) \in \mathcal{D}{\text{test}}} \mathbb{I}( \mathcal{H}(\vx; \theta_{\mathcal{H}}) = \boldsymbol{y}), \\
    \text{where\;} & \hat{\mathcal{F}} = \underset{\theta_{\hat{\mathcal{F}}}}{\operatorname{argmin}} \, \mathbb{E}_{(\vx, \mathcal{H}(\vx)) \in \mathcal{D}_T} \left[ \mathcal{L}\left(\hat{\mathcal{F}}(\vx; \theta_{\hat{\mathcal{F}}}), \mathcal{H} (\vx)\right)\right],\\
\end{aligned}
\end{equation}
where $T(\vx)$ is the operation that adds the trigger to $\vx$, $\mathbb{I}$ is the indicator function, $\boldsymbol{y}_\mathrm{target}$ is the predefined target backdoor label, $\mathcal{D}_{\mathrm{test}}$ is the victim model's test set, and $\mathcal{L}$ is a loss function (e.g., cross-entropy, or Kullback–Leibler divergence~\cite{kullback1951information}).

Defending with a honeypot layer offers several advantages: (1) The honeypot layer has a small number of parameters, resulting in minimal computational overhead for finetuning; (2) It operates solely on the output features of the victim model, avoiding the need for retraining and making it suitable for large-scale pre-trained models; (3) The backdoor is introduced exclusively into the honeypot layer, ensuring that no additional security risks are posed to the victim model.

\subsection{Finetuning the Honeypot Layer}
Intuitively, the goal defined in \Eqref{eq:2} can be achieved through a backdoor attack. By embedding a backdoor into the honeypot layer, the attacker will inevitably extract this backdoor into their substitute model when replicating the functionality of the honeypot layer. However, as noted by \citet{lv2024mea}, model extraction attacks focus primarily on the key functions of the victim model, making it difficult to extract task-irrelevant backdoors into the substitute model. Additionally, since we cannot retrain the victim model, we are unable to use functionally relevant backdoors.

Inspired by the transferability of adversarial examples~\cite{liu2017delving}, we propose using a Universal Adversarial Perturbation (UAP)~\cite{moosavi2017universal} as an effective backdoor trigger. Our objective is to find a UAP (denoted as $\boldsymbol{\delta}$) that when applied to any input image $\vx$, causes the substitute model $\hat{\mathcal{F}}$ to predict the target class $\boldsymbol{y}_{\mathrm{target}}$:
\begin{equation}
    \hat{\mathcal{F}}(T(\vx)) = \boldsymbol{y}_\mathrm{target}; \;\; \text{where} \; T(\vx) = \operatorname{clip}( \vx + \boldsymbol{\delta}, 0, 1).
\end{equation}
This perturbation $\boldsymbol{\delta}$ can be found through stochastic gradient descent (SGD)~\cite{shafahi2020universal} with a gradient sign update, similar to the Fast Gradient Sign Method (FGSM)~\cite{goodfellow2014explaining}:
\begin{equation}
\label{eq:4}
    \boldsymbol{\delta} \gets \boldsymbol{\delta} - \epsilon \cdot \operatorname{sign} (\mathbb{E}_{\vx} [\triangledown_{\boldsymbol{\delta}} \mathcal{L}(\hat{\mathcal{F}}(\vx+\boldsymbol{\delta}), \boldsymbol{y}_{\mathrm{target}})] ),
\end{equation}
where $\epsilon$ is the step size and $\operatorname{sign}(\cdot)$ is the sign function.

As the defender, obtaining the substitute model is not feasible. 
Therefore, we introduce a \textbf{\emph{shadow model}} $\mathcal{F}_{s}$ to approximate $\hat{\mathcal{F}}$ and a \textbf{\emph{shadow set}} $\mathcal{D}_{s}$ to replace the attacker's transfer set $\mathcal{D}_{T}$.
To minimize computational overhead, we use a lightweight model, such as ResNet18~\cite{he2016deep}, as $\mathcal{F}_{s}$. We employ $\mathcal{F}_{s}$ and the shadow set $\mathcal{D}_{s}$ to simulate the model extraction process. By solving \Eqref{eq:4} on $\mathcal{F}_{s}$, we obtain the UAP $\boldsymbol{\delta}$. This trigger is then used to fine-tune the Honeypot layer $\mathcal{H}$.
This process is formulated as a Bi-Level Optimization (BLO) problem:
\begin{equation}
\label{eq:5}
    \underset{\theta_{\mathcal{H}}}{\operatorname{argmin}} \, \mathbb{E}_{\vx \in \mathcal{D}_{s}}[  \underbrace{\mathcal{L}(\mathcal{H}(\vx),\mathcal{F}(\vx))}_{\text{for normal functionality}}  
      +  \underbrace{\mathcal{L}(\mathcal{H}(\vx+\boldsymbol{\delta}),\boldsymbol{y}_{\mathrm{target}})}_{\text{for backdoor injection}}],
\end{equation}

\begin{equation}
\begin{aligned}
\label{eq:5_1}
    \text{s.t.\;\;} &  \underset{\theta_{\mathcal{F}_{s}}}{\operatorname{argmin}} \, \mathbb{E}_{\vx\in \mathcal{D}_{s}} [  \mathcal{L}(\mathcal{F}_{s}(\vx), \mathcal{H}(\vx))],  \\ 
     & \underset{\boldsymbol{\delta}} {\operatorname{argmin}} \, \mathbb{E}_{\vx \in \mathcal{D}_{v}}[  \mathcal{L}(\mathcal{F}_{s}(\vx+\boldsymbol{\delta}), \boldsymbol{y}_{\mathrm{target}})],
\end{aligned}
\end{equation}
where \Eqref{eq:5} is the upper level, \Eqref{eq:5_1} is the lower level, $\mathcal{D}_{v}$ is a small verification dataset related to the victim model's task, used to verify the backdoor attack success rate, and can be collected through an online search, and the sample classes should be evenly distributed as much as possible; and $\mathcal{L}$ is the cross-entropy loss.

Specifically, the BLO framework iteratively executes the following three steps: \textbf{\emph{1) extraction simulation}}, \textbf{\emph{2) trigger generation}}, and \textbf{\emph{3) finetuning}}.
In the \textbf{extraction simulation} step, we randomly select $n$ samples from the shadow dataset $\mathcal{D}_{s}$ for each iteration. 
We query the honeypot layer $\mathcal{H}$ to obtain predictions and then use sample-prediction pairs to train the shadow model $\mathcal{F}{s}$ for $o$ epochs. 
This step simulates the process of an attacker obtaining a substitute model through model extraction.
In the \textbf{trigger generation} step, the trigger $\boldsymbol{\delta}$ is updated on $\mathcal{F}_{s}$ according to the update rule defined in \Eqref{eq:4} for $o$ epochs.
To enhance concealment, a pre-defined mask $\boldsymbol{M}$ restricts the trigger to a specific location, and momentum is incorporated for improved optimization:
\begin{equation}
\label{eq:6}
\begin{aligned}
\boldsymbol{\delta}_{i} &= \alpha \cdot \boldsymbol{\delta}_{i-1} - (1-\alpha) \cdot \epsilon \cdot \operatorname{sign} (\mathbb{E}_{\vx \in \mathcal{D}_{v}} [\boldsymbol{g}(\boldsymbol{\delta}_{i-1})]), \\
\boldsymbol{g}(\boldsymbol{\delta}) &= \triangledown_{\boldsymbol{\delta}} \mathcal{L}(\mathcal{F}_{s}(\boldsymbol{M} \odot \vx+ (1-\boldsymbol{M}) \odot \boldsymbol{\delta}), \boldsymbol{y}_{\mathrm{target}}),
\end{aligned}
\end{equation}
where $\boldsymbol{g}(\boldsymbol{\delta})$ denotes the gradient of loss $\mathcal{L}$ with respect to trigger $\boldsymbol{\delta}$, $\alpha$ is the momentum parameter and $\boldsymbol{M}$ is the binarized mask.
In the \textbf{finetuning} step, the honeypot layer is fine-tuned according to \Eqref{eq:5}, to maintain normal functionality while enhancing sensitivity to the trigger. 
This ensures that the extracted model inherits the backdoor vulnerability. 
In other words, the correlation between the UAP $\boldsymbol{\delta}$ and $y_{target}$ is fine-tuned to be a normal functionality in the honeypot layer so that it will pass on to the extracted model.
Since we cannot access the victim model's full training dataset, we also use the $n$ samples randomly selected from $\mathcal{D}_s$ in the simulation step for finetuning.
After $m$ BLO iterations, the victim model with the honeypot layer will be deployed for its API service.
We name the entire framework as \textbf{\emph{HoneypotNet}} and outline its detailed procedure in \Algref{alg1}.

\begin{algorithm}[t]
\caption{HoneypotNet}\label{alg1}
\textbf{Input}: Victim model $\mathcal{F}$, Shadow dataset $\mathcal{D}_{s}$.\\
\textbf{Output}: Honeypot Layer $\mathcal{H}$, trigger $\boldsymbol{\delta}$. 
  \begin{algorithmic}[1]
  \State Initialize $\mathcal{H}$, $\mathcal{F}_{s}$, and $\delta$.
  \For{epoch in $o$}
\State Select $i$ samples from $\mathcal{D}_s$ and query $\mathcal{H}$ 
\For{epoch in $o$} \Comment{Extraction simulation}
\State $\mathcal{L}=\sum_{\vx \in \mathcal{D}_s} \mathcal{L}^{\prime}(\mathcal{F}_{s}(\vx), \mathcal{H}(\vx) )$ 
\State $\mathcal{F}_{s} \leftarrow \operatorname{update}(\mathcal{F}_{s}, \mathcal{L})$ 
\EndFor
\For{epoch in $o$} \Comment{Trigger generation}
\State Update $\boldsymbol{\delta}$ according to \Eqref{eq:6} 
\EndFor
\For{epoch in $o$} \Comment{Finetuning}
\State Calculate the loss $\mathcal{L}$ according to \Eqref{eq:5_1}
\State $\mathcal{H} \leftarrow \operatorname{update}(\mathcal{H}, \mathcal{L})$
\EndFor
\EndFor
  \end{algorithmic}
\end{algorithm}

\subsection{Ownership Verification and Reverse Attack}
Each deployed Honeypot layer in the protected model is equipped with a unique, optimized trigger. This trigger has two main purposes: robust ownership verification and facilitating a reverse attack against any unauthorized substitute model.
\emph{\textbf{Ownership verification}}: 
To determine if a suspect model has been extracted from the protected model, the defender can query it with samples embedded with the trigger but without specifying the target class. If the suspect model shows a classification accuracy \(Acc_v\) on the target class \(y_{\text{target}}\) that exceeds a predefined threshold (e.g., 10\%), this suggests the presence of the backdoor. Such elevated accuracy provides strong evidence of model extraction and confirms ownership.
\emph{\textbf{Reverse attack}}: 
Unlike existing watermarking techniques that rely on specific watermarked samples for verification, our approach offers a more aggressive response to confirmed model extraction.
The trigger $\delta$ acts as a universal key that disrupts the functionality of the substitute model.  By embedding $\delta$ into any input sample, the defender can force the substitute model to consistently predict the target backdoor class. 
This manipulation has serious consequences for the attacker. Unaware of the backdoor, the attacker risks deploying a compromised model that results in erroneous predictions and potential reputational damage. This effectively forces the attacker to abandon their substitute model, which serves as a powerful deterrent against model extraction.


%% file: sec/4_experiment.tex
\section{Experiments}

\begin{table*}[t]
    \centering
    \setlength{\tabcolsep}{1mm}
    \begin{tabular}{l|c|ccc|ccc|ccc|ccc}
    \toprule[2pt]
    \multirow{2}{*}{\textbf{Extraction Method}}  & \multirow{2}{*}{\textbf{Model}} &  \multicolumn{3}{c|}{\textbf{CIFAR10}} & \multicolumn{3}{c|}{\textbf{CIFAR100}} & \multicolumn{3}{c|}{\textbf{CUBS200}} & \multicolumn{3}{c}{\textbf{Caltech256}}  \\
    & & $Acc_c$ & $Acc_v$ & ASR & $Acc_c$ & $Acc_v$ & ASR & $Acc_c$ & $Acc_v$ & ASR & $Acc_c$ & $Acc_v$ & ASR \\
    \midrule[1pt]
    \multirow{3}{*}{KnockoffNets} & No defense & \underline{82.96}&1.99&12.47 & \underline{52.66}&0.41&0.90 & 61.25&0.05&0.14 & \underline{73.25}&0.14&0.47 \\
     &  DVBW & 82.32&3.38&12.50 & 49.87&0.02&0.91 & \underline{64.52}&0.29&0.35 & 73.11&0.17&0.26\\
     & HoneypotNet & 82.28&\textbf{54.88}&\textbf{59.35} & 50.59&\textbf{85.61}&\textbf{85.71} & 60.04&\textbf{78.38}&\textbf{78.31} & 69.03&\textbf{79.04}&\textbf{79.13} \\
     \midrule[1pt]
    \multirow{3}{*}{ActivteThief (Entropy)} & No defense & \underline{82.08}&0.78&10.21 & 51.28&1.17&1.50 & \underline{65.21}&0.16&0.43 & 74.28&0.11&0.42 \\
    & DVBW & 81.69&1.53&9.39 & 50.52&1.98&2.36 & 61.06&0.29&0.67 & \underline{74.92}&0.41&0.93 \\
    & HoneypotNet & 82.04&\textbf{52.29}&\textbf{56.99} & \underline{52.56}&\textbf{74.09}&\textbf{74.35} & 62.53&\textbf{83.39}&\textbf{83.22} & 70.86&\textbf{77.29}&\textbf{77.43} \\
    \midrule[1pt]
    \multirow{3}{*}{ActivteThief (k-Center)} & No defense & \underline{83.31}& 1.03&11.38 & \underline{52.23}&0.17&0.89 & \underline{66.17}&0.03&0.26 & \underline{76.83}&0.15&0.34 \\
    & DVBW & 82.20& 3.90&10.32 & \underline{52.23}&0.31&0.54 & 63.19&0.39&0.44 & 74.06&0.23&0.77 \\
    & HoneypotNet & 82.13&\textbf{64.00}&\textbf{67.49} & 52.11&\textbf{74.48}&\textbf{74.63} & 65.19&\textbf{80.20}&\textbf{80.27} & 72.57&\textbf{80.62}&\textbf{80.80} \\
    \midrule[1pt]
    \multirow{3}{*}{SPSG} & No defense & \underline{85.47}&0.36&10.31 & \underline{53.12}&1.23&1.37 & \underline{63.18}&0.12&0.42 & \underline{71.74}&0.09&0.41 \\
    & DVBW & 83.13&1.79&12.20 & 51.18&0.77&0.97 & 60.42&0.77&0.89 & 69.91&0.82&0.95 \\
    & HoneypotNet & 83.33&\textbf{62.93}&\textbf{66.12} & 52.05&\textbf{76.92}&\textbf{77.11} & 61.37&\textbf{83.56}&\textbf{83.51} & 67.06&\textbf{77.91}&\textbf{77.88} \\
    \midrule[1pt]
    \multirow{3}{*}{BlackBox Dissector} & No defense & \underline{76.64}&0.42&9.47 & \underline{40.02}&0.38&0.60 & \underline{33.04}&0.24&0.76 & \underline{50.50}&0.07&0.28 \\
    & DVBW & 74.58&2.14&12.62 & 37.15&0.12&0.49 & 31.83&0.16&0.69 & 45.75&0.43&0.77 \\
    & HoneypotNet & 74.97&\textbf{76.26}&\textbf{78.59} & 38.81&\textbf{79.87}&\textbf{80.05} & 30.50&\textbf{92.61}&\textbf{92.35} & 47.95&\textbf{78.90}&\textbf{78.98} \\
    \bottomrule[2pt]
    \end{tabular}
    \caption{Effectiveness of HonetpotNet: the $Acc_c$ $\uparrow$ (\%), $Acc_v$ $\uparrow$ (\%), and ASR $\uparrow$ (\%) of extracted substitute model from different defense methods by five model extraction attacks under 30k queries.
    }
    \label{tab:eff}
\end{table*}

\subsection{Experimental Setup}

\textbf{Victim and Shadow Models} \quad
Following previous works \cite{orekondy19knockoff,orekondy2019prediction}, the victim models we consider are ResNet34~\cite{he2016deep} models trained on four datasets: CIFAR10, CIFAR100~\cite{krizhevsky2009learning}, Caltech256~\cite{griffin2007caltech}, and CUBS200~\cite{wah2011caltech}. The clean test accuracy of victim models are $91.56\%$, $71.57\%$, $77.11\%$, and $78.44\%$. 
Consistent with prior work~\cite{orekondy19knockoff,juuti2019prada,pal2020activethief}, we initially assume the attacker employs the same architecture to train the substitute model. 
The impact of varying substitute architectures is also explored in our analysis.
For the shadow model, we opt for a smaller model, i.e., ResNet18~\cite{he2016deep}, to minimize computational overhead.

\textbf{Attack and Shadow Datasets} \quad
We chose the ImageNet~\cite{russakovsky2015imagenet} dataset as the attack dataset, which contains 1.2M images. 
We resize images to fit the input size of victim models.
For the shadow dataset, we randomly select 5,000 images from the CC3M~\cite{sharma2018conceptual} dataset due to its distinct distribution from ImageNet. 
This simulates a realistic scenario where the defender does not know what the attacker will use as the attack dataset.

\textbf{Training and Extraction Configuration} \quad
We perform BLO for 30 iterations, with each iteration comprising three steps: (1) Extraction Simulation: We train a ResNet18 for 5 epochs using SGD (momentum 0.9, learning rate 0.1, cosine annealing) on a transfer set generated by querying the honeypot layer. (2) Trigger Generation: We update the trigger for 5 epochs with momentum 0.9. (3) Finetuning: We fine-tune the honeypot layer for 5 epochs using SGD (momentum 0.9, learning rate 0.02, cosine annealing).
For models with a small input image size (CIFAR10 and CIFAR100), we select a $6\times6$ square located 4 pixels away from the upper-left corner as the trigger location. 
For models with a larger input image size (Caltech256 and CUBS200), we choose a $28\times28$ square trigger at the same location. For simplicity, the last class is designated as the target class.
Following previous work \cite{orekondy19knockoff,orekondy2019prediction,pal2020activethief}, we train substitute models for 200 epochs using SGD (momentum 0.9, learning rate 0.02, cosine annealing).

\textbf{Evaluation Metrics} \quad
We employ three metrics: Clean Test Accuracy ($Acc_c$), Verification Test Accuracy ($Acc_v$), and Attack Success Rate (ASR).  
$Acc_c$ measures substitute models' performance on clean test samples, reflecting defense's ability to preserve victim model utility while remaining undetectable to attackers.
$Acc_v$ assesses substitute models' accuracy on a set of triggered samples from the non-target classes on the target label. 
A high $Acc_v$ indicates a successful ownership verification. 
ASR quantifies the success rate of defender in reverse attack substitute models by forcing it to predict the target label on any triggered input.

\textbf{Extraction Methods and Baseline Defenses} \quad
To evaluate the effectiveness of HoneypotNet, we apply five state-of-the-art extraction attacks, i.e., KnockoffNets~\cite{orekondy19knockoff}, ActiveThief (Entropy \& k-Center)~\cite{pal2020activethief}, SPSG~\cite{zhao2024fully}, and BlackBox Dissector~\cite{wang2021black}.
Notably, BlackBox Dissector method performs extraction under a hard-label setting, posing a unique challenge to our defense.
We compare HoneypotNet with two baseline methods: no defense and DVBW\cite{li2023black}, a defense method employing backdoor attacks for dataset ownership verification.

\subsection{Main Results}

\textbf{Effectiveness of HonetpotNet} \quad
Table \ref{tab:eff} presents the results with a query budget of 30,000.
The attack is successful, as all five extracted substitute models exhibit a high $Acc_c$ value.
This proves that HoneypotNet does not harm the utility of models and does not alert the attacker.
Compared to the negligible $Acc_v$ values of undefended models and those protected by DVBW \cite{li2023black}, HoneypotNet achieves significantly higher verification accuracy (52.29\%-92.61\%), indicating its effectiveness in verifying model ownership.
Most importantly, HoneypotNet achieves consistently high ASR, ranging from 56.99\% to 92.35\%, demonstrating its ability to effectively inject backdoors into substitute models and enable powerful reverse attacks. 
Notably, our method proves remarkably effective even in a challenging scenario of hard labels, achieving a 78.59\%-92.35\% ASR on substitute models extracted by BlackBox Dissector. This underscores its practical utility.

\begin{figure}[t]
    \centering
    \includegraphics[width=0.7\columnwidth]{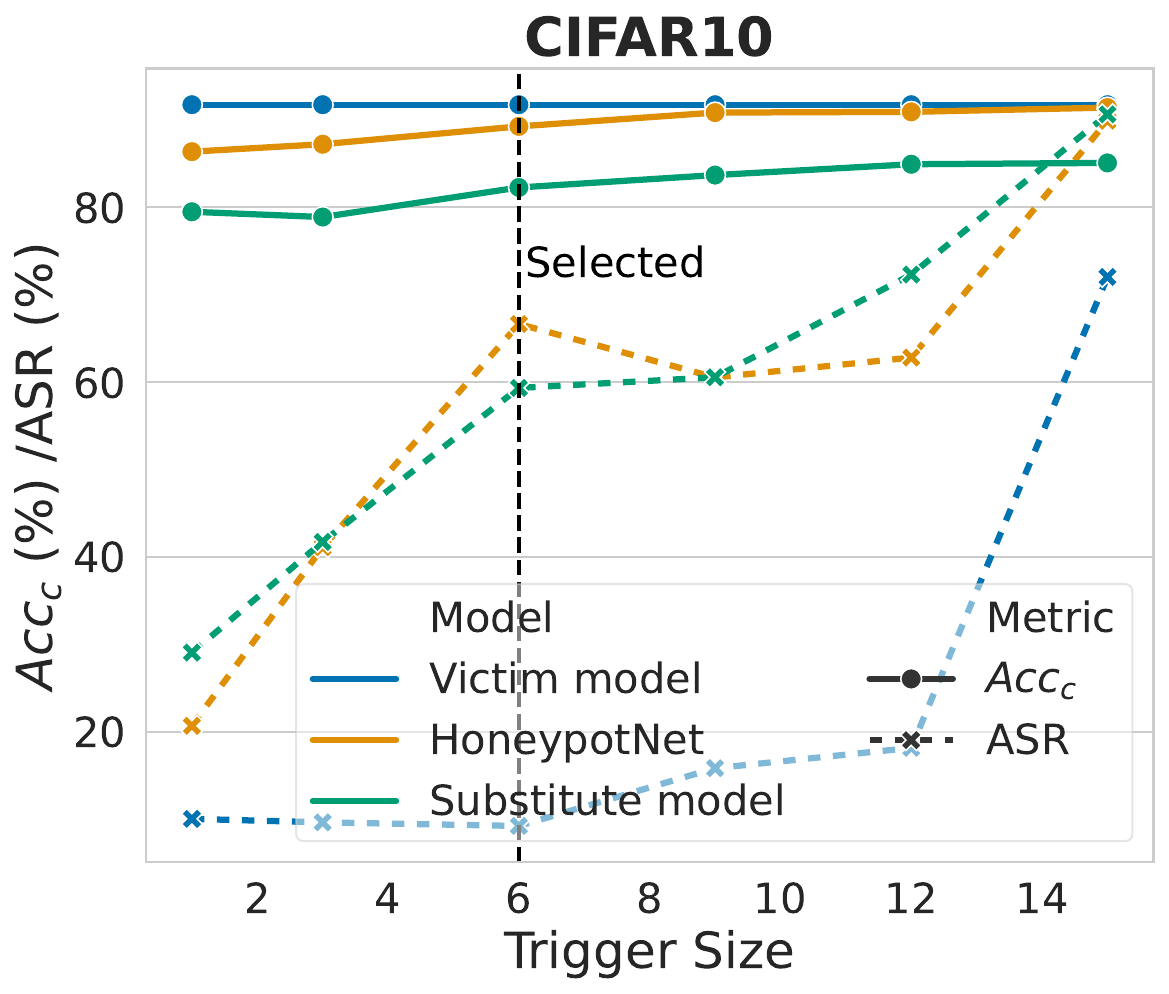}
    \caption{The impact of trigger size on the victim model, HoneypotNet, and the substitute model ( extracted by the KnockoffNets attack under 30k queries) on CIFAR10.}
    \label{fig:trigger}
\end{figure}

\textbf{Influence of Trigger Size}  \quad
Here, we test the impact of trigger size on the results with the KnockoffNets and CIFAR10.
We test trigger sizes varying from $1\times1$ to $15\times15$ and report $Acc_c$ and ASR of the victim, HoneypotNet, and substitute models in \Figref{fig:trigger}.
The trend suggests that as the trigger size increases, the ASR becomes higher, indicating a better attack effect. It is worth noting that a larger trigger will also impact the victim model, showing an abrupt increase in ASR when the trigger size is larger than 6. Interestingly, we find that the clean performance ($Acc_c$) of HoneypotNet increases with the increase of the trigger size.
We believe this is because larger triggers have a stronger attack capability and thus are easier to learn without losing much of the original performance.
Overall, this experiment suggests that one should balance the trigger size and protection effect in real-world scenarios to strike a better trade-off. 

\begin{figure*}[t]
    \centering
    \includegraphics[width=\textwidth]{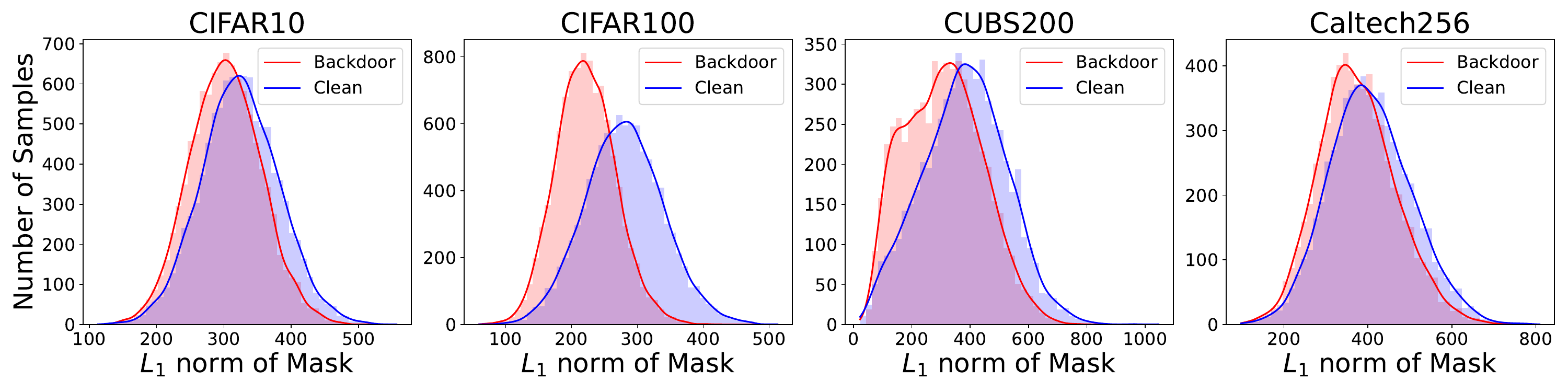}
    \caption{
    The $L_1$ norm distributions of the detected backdoor patterns for clean and backdoor samples by the backdoor detection method Cognitive Distillation (CD) \cite{huang2023distilling}.
    }
    \label{fig:cd}
\end{figure*}

\begin{table}[t]\small
    \centering
    \setlength{\tabcolsep}{1mm}
    \begin{tabular}{l|cccc}
    \toprule[2pt]
    \textbf{Arc.}  & \textbf{CIFAR10} & \textbf{CIFAR100} & \textbf{CUBS200} & \textbf{Caltech256}  \\
    \midrule[1pt]
    ResNet34 & {59.35} & 85.71 & 78.31 & \textbf{79.13}  \\
    ResNet18 & {59.17} & 69.55 & 77.55 & 66.42   \\
    ResNet50 & {59.53} & 67.35 & 67.63 & 60.80   \\
    VGG16 & \textbf{{97.16}} & \textbf{87.10} & \textbf{89.82} & 62.17  \\
    DenseNet121 & {51.68} & 53.72 & 65.46 & 58.00 \\
    \bottomrule[2pt]
    \end{tabular}
    \caption{The influence of different substitute model architectures on ASR (\%). All substitute models were extracted by the KnockoffNets attack under 30k queries.}
    \label{tab:arc}
\end{table}

\textbf{Influence of Substitute Model Architecture} \quad
Contrary to the previous assumption that the victim and substitute models have same architectures, here we test more substitute architectures that are different from the victim. We take the KnockoffNets attack as an example and report the results in Table \ref{tab:arc}.
Given that our backdoor trigger is based on the transferability of UAP, its effectiveness can vary depending on the model architecture, as adversarial transferability is sensitive to model architecture.
For instance, the VGG16 model exhibits lower robustness, leading to higher attack success rates on various datasets, such as reaching 97.16\% ASR on CIFAR10.
The DenseNet121 model on the other hand has a slightly lower ASR. 
However, the lowest ASR achieved by our HoneypotNet is still above 51\%, demonstrating its effectiveness even when extracted using different substitute models. 
Additionally, we believe that using a more complex shadow model or even an ensemble of multiple shadow models can further improve HoneypotNet.

\textbf{Evading Potential Backdoor Detection} \quad
The attacker may leverage a backdoor detection method to detect whether there is a backdoor in the substitute model. 
To test this, here we consider a state-of-the-art backdoor detection method Cognitive Distillation (CD)~\cite{huang2023distilling} to test where it can detect the backdoor trigger pattern from the substitute model (extracted by KnockoffNets on CIFAR10).
CD identifies backdoors by extracting minimal backdoor patterns (cognitive patterns) for a test image and comparing their $L_1$ norms between clean and potentially backdoored samples. 
A lower $L_1$ norm in backdoor samples suggests the presence of a shortcut pattern, indicating a backdoor.
As shown in \Figref{fig:cd}, the $L_1$ norm distributions of clean vs. backdoor samples are very similar. 
This can be attributed to the inherent nature of our UAP-based trigger, which seamlessly integrates with the model's decision boundaries, mimicking natural features and evading detection by methods like CD that rely on identifying anomalous patterns.

\begin{figure}[t]
    \centering
    \includegraphics[width=\columnwidth]{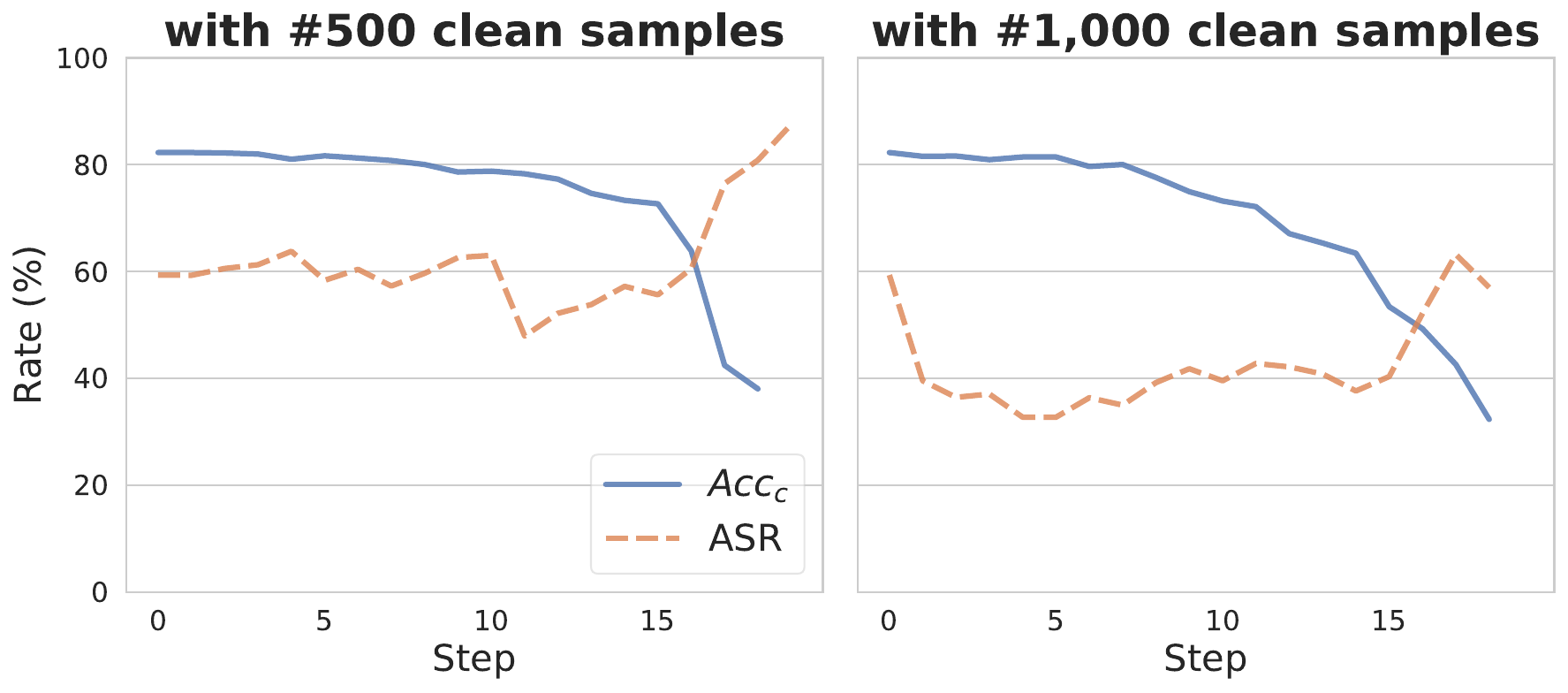}
    \caption{Robustness of HoneypotNet against Reconstructive Neuron Pruning (RNP) on the CIFAR10 dataset. 
    The $Acc_c$ and ASR are reported at different steps of the RNP process with 500 (left) and 1,000 (right) clean samples.}
    \label{fig:rnp}
\end{figure}

\textbf{Robustness against Backdoor Neuron Pruning} \quad
To assess the robustness of our injected backdoor against possible neuron pruning by attackers, we evaluate the performance of HoneypotNet when countered by the Reconstructive Neuron Pruning (RNP) method~\cite{li2023reconstructive}. 
RNP aims to identify and prune backdoor neurons using only a small set of clean samples by leveraging an asymmetric unlearning-recovering process.
We conduct experiments on the CIFAR10 dataset, utilizing the extracted substitute model from the KnockoffNets under 30k queries. 
We vary the number of clean samples used by RNP for defense and report the $Acc_c$ and ASR at different steps of the RNP process in the \Figref{fig:rnp}. 
Our results demonstrate that the ASR remains consistently high, with minimal impact from the varying defense data size. 
This highlights the robust nature of our backdoor injection, which seamlessly integrates the trigger into the model's normal functionality, making it difficult to be detected and pruned by defense mechanisms like RNP. This finding further underscores the effectiveness of HoneypotNet against sophisticated pruning-based defenses.

%% file: sec/5_conclusion.tex
\section{Conclusion}
We introduce a novel defense paradigm called \emph{attack as defense} which actively releases poisonous outputs to counteract and disrupt model extraction attacks. Unlike existing defense methods, \emph{attack as defense} not only verifies model ownership but also hampers the functionality of substitute models. 
We present a specific implementation of this paradigm, \textbf{HoneypotNet}, which replaces the classification head of the victim model with a \emph{honeypot layer} designed to generate both correct and poisonous probability vectors. The honeypot layer is fine-tuned using a shadow model and shadow dataset through bi-level optimization (BLO).
Our empirical evaluation of HoneypotNet across four benchmark datasets, five model extraction attacks, and various substitute model architectures demonstrates its effectiveness. We hope our work will inspire further research into \emph{attack as defense} strategies against model extraction attacks.

\section*{Acknowledgments}
This work is in part supported by the National Key R\&D Program of China (Grant No. 2022ZD0160103), the National Natural Science Foundation of China (Grant No. 62276067), and Shanghai Artificial Intelligence Laboratory.